\documentclass[aps,scriptaddress,twocolumn,prl,showpacs]{revtex4}

	\usepackage{amssymb} 
	\usepackage{amsmath}
	\usepackage{slashed}
	\usepackage{makeidx}
	\usepackage{amsfonts}
	\usepackage[ansinew]{inputenc}
	\usepackage[usenames,dvipsnames]{pstricks}
\usepackage{graphicx}  
\usepackage{dcolumn}   
\usepackage{bm}        
\usepackage{amssymb}   
	\usepackage{subfigure}
	\usepackage{epsfig}
	\usepackage{pst-grad} 
	\usepackage{pst-plot} 
	\usepackage[colorlinks,hyperindex]{hyperref}
	\hypersetup
	{
		colorlinks,%
		citecolor=black,%
		linkcolor=black,%
		urlcolor=black,%
	}

\makeindex

\begin{document}

\title{Spontaneous supersymmetry breaking in the $2d$ $\mathcal{N}=1$ Wess-Zumino model}

\author{Kyle Steinhauer and Urs Wenger}
\email{wenger@itp.unibe.ch}
\affiliation{Albert Einstein Center for Fundamental Physics, University of Bern,
Sidlerstrasse 5, 3012 Bern, Switzerland.}

\date{\today}

\begin{abstract}
  We study the phase diagram of the two-dimensional $\mathcal{N}=1$
  Wess-Zumino model on the lattice using Wilson fermions and the
  fermion loop formulation. We give a complete nonperturbative
  determination of the ground state structure in the continuum and
  infinite volume limit. We also present a determination of the
  particle spectrum in the supersymmetric phase, in the supersymmetry
  broken phase and across the supersymmetry breaking phase
  transition. In the supersymmetry broken phase we observe the
  emergence of the Goldstino particle.
\end{abstract}

\pacs{11.30.Pb,11.30.Qc,12.60.Jv,05.50.+q}

\maketitle

\section{Introduction}

Understanding the spontaneous breakdown of supersymmetry is a generic
nonperturbative problem which is relevant not only for particle
physics but in fact for many physical systems beyond quantum field
theories. The $\mathcal{N} = 1$ Wess-Zumino model
\cite{Wess:1973kz,Ferrara:1975nf} in two dimensions is one of the
simplest supersymmetric quantum field theories which allows for
spontaneous supersymmetry breaking since it enjoys the necessary but
not sufficient condition of a vanishing Witten index
\cite{Witten:1982df}.  The model has been ana\-lysed employing various
approaches such as Monte Carlo methods
\cite{BartlesBronzan,Ranft:1983ag}, Hamiltonian techniques
\cite{Elitzur:1982vh,Beccaria:2004pa,Beccaria:2004ds}, or exact
renormalisation group methods \cite{Synatschke:2009nm}. Wilson
derivatives for fermions and bosons, guaranteeing a supersymmetric
continuum limit \cite{Golterman:1988ta}, were used in
\cite{Catterall:2003ae} and a numerical analysis of the phase diagram
using the SLAC derivative has been conducted in
\cite{Wozar:2011gu}. All approaches use various regulators which are
more or less difficult to control. In this letter we report on our
results for the two-dimensional $\mathcal{N}=1$ Wess-Zumino model
regularized on a Euclidean spacetime lattice. The discretization using
Wilson derivatives for the fermions and bosons \cite{Golterman:1988ta}
together with the fermion loop formulation and a novel algorithm
\cite{Wenger:2008tq} allows to systematically remove all effects from
the IR and UV regulators by explicitly taking the necessary limits in
a completely controlled way. One reason why this has not been achieved
so far with other methods is the fact that all supersymmetric systems
with spontaneously broken supersymmetry suffer from a fermion sign
problem related to the vanishing of the Witten index
\cite{Baumgartner:2011cm}. However, that sign problem can be
circumvented in our approach by using the exact reformulation of the
lattice model in terms of fermion loops \cite{Baumgartner:2011cm}. In
this formulation the partition function is obtained as a sum over
closed fermion loop configurations and separates naturally into its
bosonic and fermionic parts for which the sign is perfectly under
control.  Efficient simulations with an open fermion string (or
fermionic worm) algorithm \cite{Wenger:2008tq} are then possible even
in the phase with spontaneously broken supersymmetry where the
massless Goldstino mode is present.

The two-dimensional $\mathcal N=1$ Wess-Zumino model
\cite{Wess:1973kz,Ferrara:1975nf} contains a real two component
Majorana spinor $\psi$ and a real bosonic field $\phi$ and is
described in Euclidean spacetime by the on-shell continuum action
\begin{equation}
\label{mainWZcontinuumaction}
S=\int d^2x \, \left\{ \frac{1}{2} (\partial_{\mu} \phi)^2 +\frac{1}{2} \overline{\psi}D \psi +\frac{\left[P'(\phi)\right]^2}{2}  \right\} 
\end{equation} 
where $D=\left[\slashed{\partial}+P''(\phi)\right]$ is the Majorana
Dirac operator.  Here, $P(\phi)$ denotes a generic superpotential and
$P'$ and $P''$ its first and second derivative with respect to $\phi$,
respectively.  The action is invariant under a supersymmetry
transformation $\delta$ which transforms $\phi$, $\psi$ and
$\overline{\psi}$ as
\begin{equation}
\label{SUSYsymmetry_WZ}
 \delta \phi = \overline{\epsilon}\psi , \quad \delta \psi = (\slashed{\partial}\phi-P')\epsilon, \quad \delta\overline{\psi}=0,
\end{equation} 
where $\epsilon$ is a constant Majorana spinor. In the following we
will concentrate on the specific superpotential
 \begin{equation}
\label{superpotentialWZ}
 P(\phi)=\frac{1}{3}g\phi^3-\frac{m^2}{4g}\phi \, .
\end{equation} 
With this potential, the action is also invariant under a discrete
$\mathbb{Z}_2$/chiral symmetry transformation
\begin{equation}
\phi \rightarrow - \phi, \quad \psi \rightarrow \sigma_3 \psi , \quad
\overline{\psi} \rightarrow - \overline{\psi} \sigma_3
\end{equation}
which in the following we denote by $\mathbb{Z}_2^\chi$ symmetry. The
potential yields a vanishing Witten index $W = 0$ and hence allows for
spontaneous supersymmetry breaking \cite{Witten:1982df}. This can for
example be derived from the transformation properties of the Pfaffian
$\text{Pf}(D)$ under the $\mathbb{Z}_2$ symmetry $\phi\to -\phi$
\cite{Baumgartner:2012np}.

\section{Fermion Loop Formulation}
When the model is regularized on a discrete spacetime lattice both the
$\mathbb{Z}_2^\chi$ and the supersymmetry are broken explicitly, but
the discretization can be chosen such that the restoration of the
symmetries is guaranteed in the continuum limit
\cite{Golterman:1988ta}. This can be achieved because the model is
superrenormalisable and only one counterterm is necessary to
renormalize the bare mass $m$, while the coupling $g$ is not
renormalized and can hence be used to define the continuum limit $ag
\rightarrow 0$ where $a$ is the lattice spacing.  The loop formulation
is obtained by constructing an exact hopping expansion of the fermion
action to all orders. When expanding the Boltzmann factor and
subsequently performing the integration over the fermion fields, the
nilpotency of the Grassmann elements ensures that only closed,
nonoriented and selfavoiding fermion loops survive.  The partition
function then becomes a sum over all fermion loop configurations $l
\in \mathcal{L}$,
\begin{equation}
 Z_\mathcal{L}=\sum_{l \in \mathcal{L}} \prod_x w_l(x)\,,
\label{eq:ZL}
\end{equation} 
where the weight for a given loop configuration is a product over site
weights $w_l(x)$ which depend only on the local geometry of the loop
at the lattice site $x$, if a fermion loop is present, or on an
integral over an ultralocal function of the bosonic field $\phi$.  The
configuration space of all loop configurations $\mathcal{L}$ naturally
separates into equivalence classes $\mathcal{L}_{ij}$ characterized by
the even or odd number of loops winding around the lattice in the
spatial and temporal direction, respectively. The loop configurations
in each equivalence class pick up a definite sign depending on the
chosen boundary conditions (b.c.) for the Majorana fermion
\cite{Wolff:2007ip}, so the partition function in eq.(\ref{eq:ZL})
represents a system with unspecified (or fluctuating) b.c., while the
one with periodic b.c.,
\begin{equation}
 W \propto Z_{pp} = Z_{\mathcal{L}_{00}}-Z_{\mathcal{L}_{10}}-Z_{\mathcal{L}_{01}}-Z_{\mathcal{L}_{11}} \, ,
\label{eq:Zpp}
\end{equation} 
is proportional to the Witten index and the one with antiperiodic
b.c.~in time,
\begin{equation}
Z_{ap} = Z_{\mathcal{L}_{00}} +
Z_{\mathcal{L}_{10}}-Z_{\mathcal{L}_{01}} + Z_{\mathcal{L}_{11}} \, ,
\label{eq:Zap}
\end{equation}
describes the system at finite temperature.  Note that the weight is
not necessarily positive definite in each of the sectors, but
sufficiently close to the continuum limit it turns out to be so.  As
described in \cite{Wenger:2008tq} the system can most efficiently be
simulated, essentially without critical slowing down, by introducing
an open fermion string corresponding to the insertion of a Majorana
fermion pair. By letting the ends of the string move around the
lattice by a standard Metropolis update procedure, one samples the
fermion 2-point function as well as the relative weights between
$Z_{\mathcal{L}_{00}}$, $Z_{\mathcal{L}_{10}}$, $Z_{\mathcal{L}_{01}}$
and $Z_{\mathcal{L}_{11}}$ which allows precise determinations of
eqs.(\ref{eq:Zpp}) and (\ref{eq:Zap}) {\it a posteriori}.  Finally,
the bosonic fields are integrated over by standard Monte Carlo methods
using a Metropolis algorithm.

\section{Vacuum Structure}

The vacuum structure of the system depends on the two bare parameters
$m$ and $g$ and is hence a function of the dimensionless ratio $f
\equiv g/m$. The expected symmetry breaking pattern in the continuum
\cite{Witten:1982df} is characterized by a supersymmetric phase with
spontaneously broken $\mathbb{Z}_2^\chi$ symmetry and a unique
(bosonic or fermionic) vacuum (groundstate) at small $f$, and a
$\mathbb{Z}_2^\chi$ symmetric phase at large $f$ with spontaneously
broken supersymmetry accompanied by tunnelling between the bosonic or
fermionic vacua (groundstates).  The two phases are separated by a
phase transition at $f_c=g/m_c$.

\begin{figure}[!tb]
\begin{center}
\includegraphics[scale=0.8]{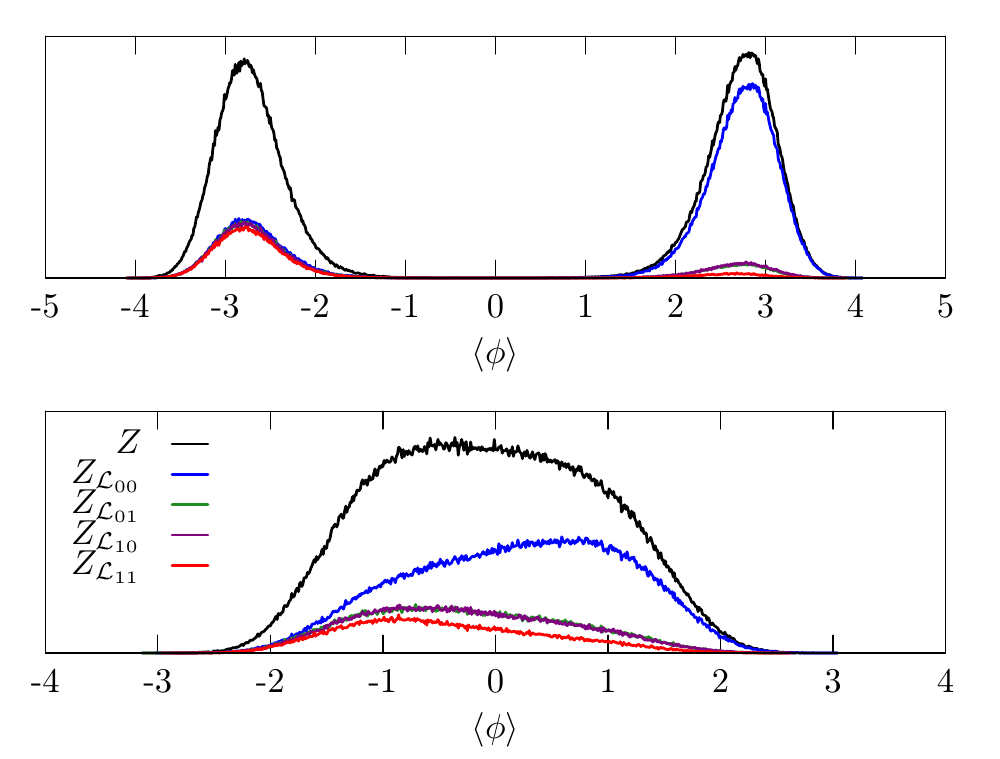}
\caption{Histograms of partition functions $Z_{\mathcal{L}_{ij}}$ on a
  $8 \times 8$ lattice at fixed lattice spacing $ag=0.0625$ for two
  different couplings $f<f_c$ (top) and $f>f_c$ (bottom).}
\label{fig_phibehaviour_L8x8Histogram_g0koma125}
\end{center}
\end{figure}

In FIG.~\ref{fig_phibehaviour_L8x8Histogram_g0koma125} we show
histograms of the partition functions $Z_{\mathcal{L}_{ij}}$ as a
function of the vacuum expectation value $\langle \phi \rangle$ of the
bosonic field in both phases.
The top panel shows data for $f<f_c$ ($\mathbb{Z}_2^\chi$
broken/supersymmetric) where $\langle \phi \rangle=\pm m/(2g)$
corresponds to the two classical minima of the potential with
$Z_{pp}/Z_{ap}=\pm 1$ in the continuum. From the plots we infer that
the groundstate at $+m/2g$, where $Z\simeq Z_{\mathcal{L}_{00}}$ and
$Z_{\mathcal{L}_{10}} \simeq Z_{\mathcal{L}_{01}} \simeq
Z_{\mathcal{L}_{11}} \simeq 0$, and hence $Z_{pp}/Z_{ap} \simeq +1$,
corresponds to the bosonic vacuum while the groundstate at $-m/2g$,
where $Z_{\mathcal{L}_{00}}\simeq Z_{\mathcal{L}_{10}} \simeq
Z_{\mathcal{L}_{01}} \simeq Z_{\mathcal{L}_{11}} \simeq Z/4$, and
hence $Z_{pp} \simeq -Z_{ap}$, corresponds to the fermionic one \cite{Baumgartner:2011jw}. In
the infinite volume limit either the bosonic or fermionic groundstate
is selected and consequently supersymmetry is intact (but the
$\mathbb{Z}_2^\chi$ symmetry is spontaneously broken).  For increasing
$f$ the tunnelling between the groundstates is enhanced and eventually
triggers the restoration of the $\mathbb{Z}_2^\chi$ symmetry
accompanied by the spontaneous breakdown of the supersymmetry at
$f_c$. This is illustrated in the lower panel of
FIG.~\ref{fig_phibehaviour_L8x8Histogram_g0koma125} where the data for
$f>f_c$ clearly displays $\langle \phi \rangle \simeq 0$ and
$Z_{pp}/Z_{ap} \simeq 0$, both of which become exactly zero in the
continuum limit \cite{Baumgartner:2011jw}. Note that the skewness of the distribution is due to
the residual $\mathbb{Z}_2^\chi$ symmetry breaking at finite lattice
spacing. Whether the spontaneous phase transition survives the
continuum and infinite volume limit, i.e., whether $f_c$ remains
finite and nonzero, needs to be investigated by quantitatively
determining $f_c$ at various lattice spacings and volumes and
carefully taking first the infinite volume limit followed by the
continuum one.

The (pseudo-)critical point $am_c$ of the spontaneous
$\mathbb{Z}_2^\chi$ symmetry breaking phase transition is determined
at fixed lattice spacing $a g$ by considering the intersection point
of the Binder cumulant $U=1-\langle\phi^4\rangle/3\langle
\phi^2\rangle^2$ obtained from different volumes using $Z_{ap}$
\cite{Baumgartner:2011jw}.
\begin{figure}[tb]
\begin{center}
\includegraphics[scale=0.355]{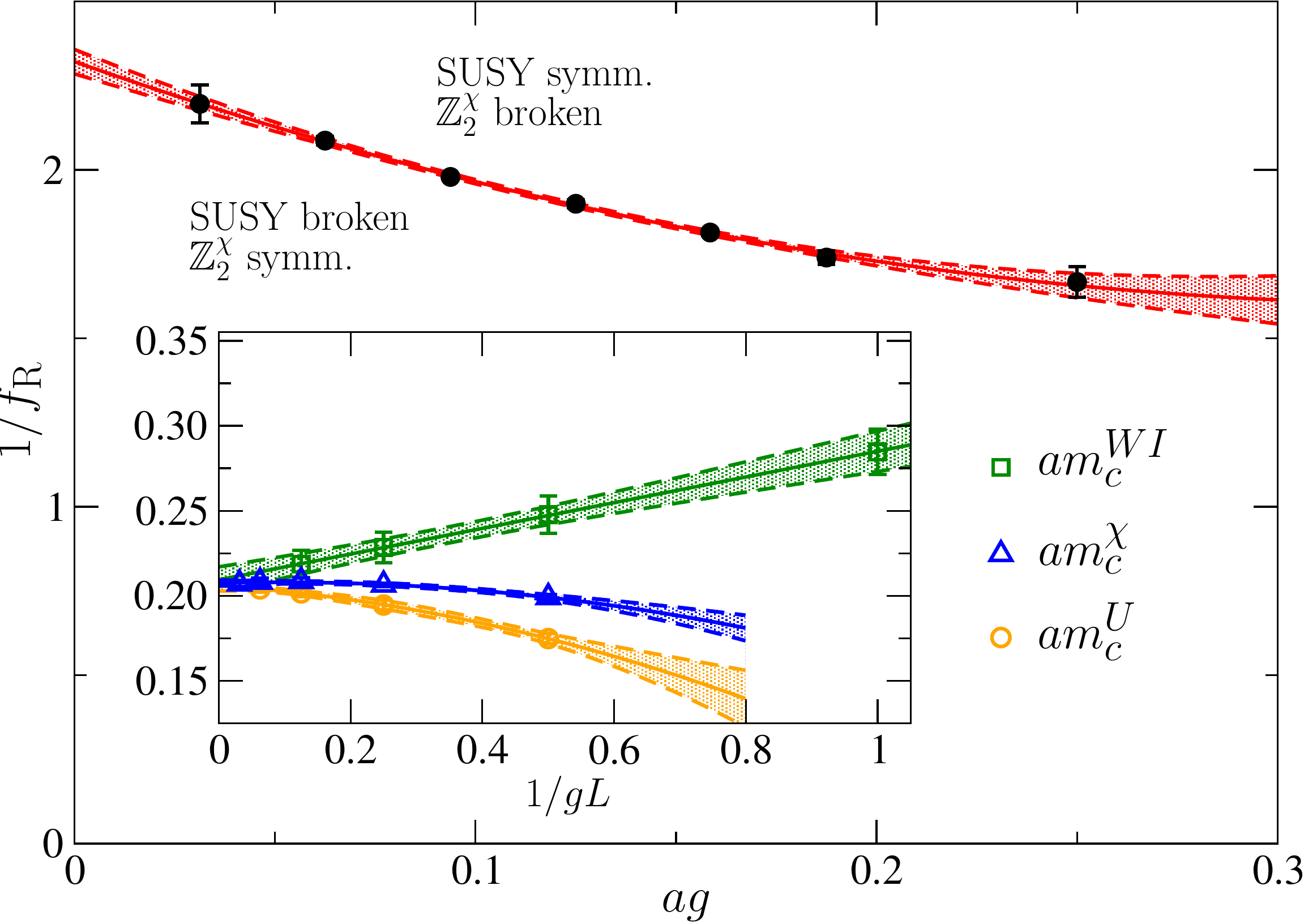}
\caption{
Continuum limit of the inverse critical coupling $1/f^R_c$.
The inset shows the infinite volume extrapolations
  of $m_c$ from various definitions at $a g=0.125$.}
\label{fig_renormalisedphasediag}
\end{center}
\end{figure}
This can be compared with the determination from the peak of the
susceptibility $\chi$ of the average sign of the bosonic field.  In
the inset of FIG.~\ref{fig_renormalisedphasediag} we show data
exemplarily for $ag=0.125$ extrapolated to the infinite volume limit
$gL \rightarrow \infty$ using linear and quadratic terms in
$1/gL$. Both observables yield values which agree in the thermodynamic
limit.  Similarly, the (pseudo-)critical point of the spontaneous
supersymmetry breaking phase transition can be determined from the
supersymmetric Ward identity $\langle P'/m \rangle$ yielding results
which in the infinite volume limit are in agreement with the
determinations from the $\mathbb{Z}_2^\chi$ transition already at
finite lattice spacing. In order to take the continuum limit this
procedure is repeated for a range of lattice spacings and the
resulting bare critical couplings $f_c=g/m_c(ag)$ are renormalised by
subtracting the logarithmically divergent one-loop self energy from
the bare mass $m^2$ and computing the renormalised critical coupling
$f^R_c=g/m^R_c(ag)$.  In FIG.~\ref{fig_renormalisedphasediag} we show
$1/f^R_c$ in the infinite volume limit as a function of the lattice
spacing together with an extrapolation to the continuum using
corrections linear plus quadratic in $a$.  The average of this
extrapolation with one using only a linear correction yields
\begin{equation}
 1/f^R_c=2.286(28)(36)\, 
\end{equation} 
where the first error is statistical and the second comes from the
difference of the two extrapolations. The result demonstrates that the
supersymmetry breaking phase transition coinciding with the
$\mathbb{Z}_2^\chi$ symmetry restoration survives the infinite volume
and the continuum limit, and it provides a precise nonperturbative
determination of the perturbatively renormalised critical coupling.
Finally we note that the determination in \cite{Wozar:2011gu} using
SLAC fermions is fully compatible with our result once the exact same
renormalization procedure is applied.

\section{Mass Spectrum}
\begin{figure}[!tb]
\begin{center}
\includegraphics[scale=0.35]{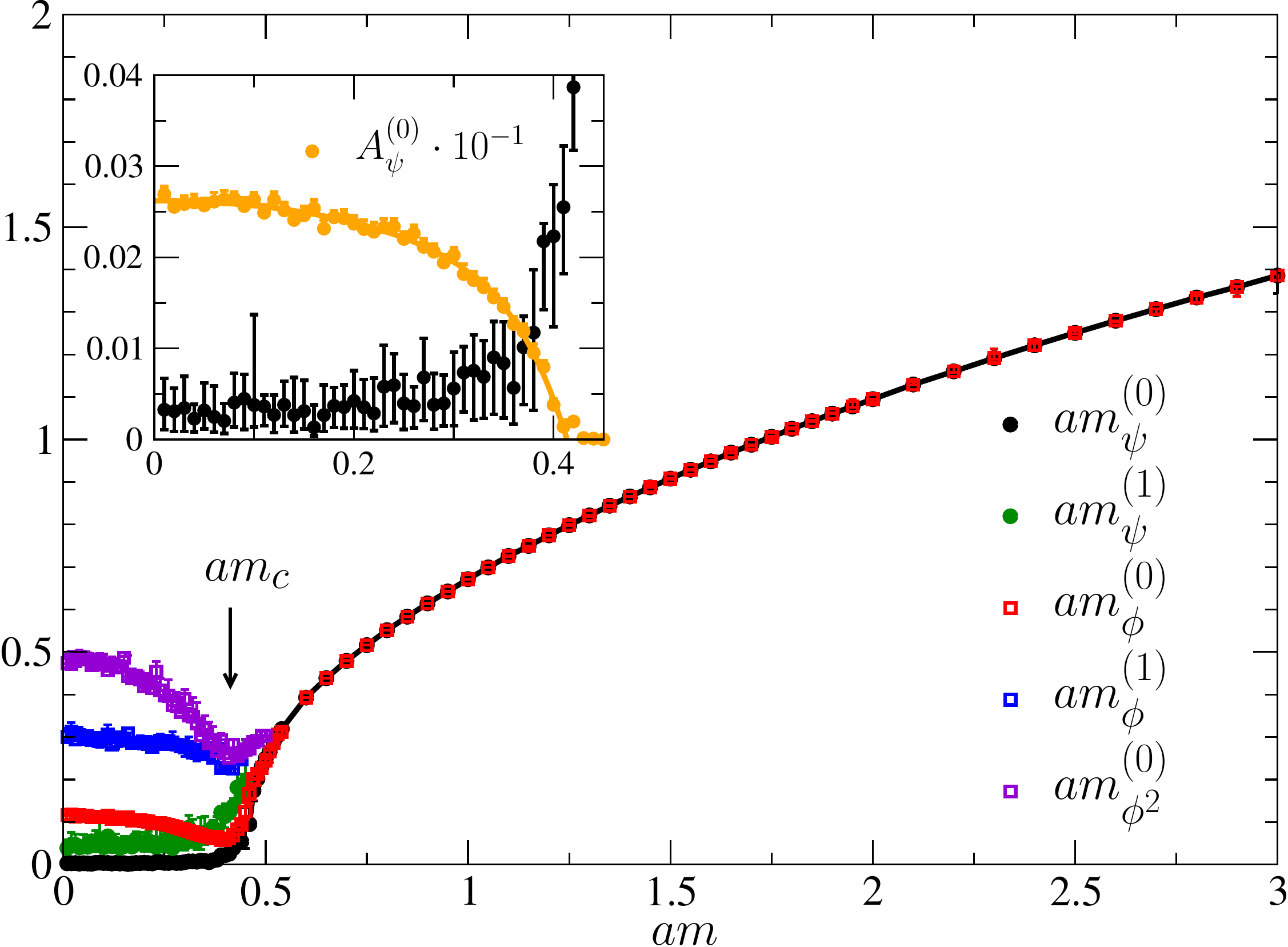}
\caption{The lowest boson and fermion masses at $a g=0.25$ in the
  bosonic sector $Z_{\mathcal{L}_{00}}$.  The inset shows a zoom of
  the Goldstino mass $m_{\psi}^{(0)}$ and its amplitude
  $A_{\psi}^{(0)}$.  }
\label{fig_massdegeneracyfinitespacing}
\end{center}
\end{figure}

Next we consider the mass spectrum of the system below, above and
across the phase transition. The lowest masses are obtained from the
exponential temporal decay of 2-point correlation functions $\langle
\mathcal{O}(t) \mathcal{O}(0)\rangle$ of appropriate fermionic or
bosonic operators $\mathcal{O}$ projected to zero spatial momentum
\cite{Baumgartner:2013ara}. Due to the open fermion string algorithm
the correlation functions can be determined to very high accuracy even
in the massless phase or when the signal falls off by many orders of
magnitude \cite{Wenger:2008tq}.  In
FIG.~\ref{fig_massdegeneracyfinitespacing} we show the lowest boson
and fermion masses $m_{\phi}^{(0)},m_{\phi}^{(1)}$ and
$m_{\psi}^{(0)},m_{\psi}^{(1)}$, respectively, in the bosonic sector
$Z_{\mathcal{L}_{00}}$ at $ag=0.25$ on a lattice with extent
$128\times 48$.
For bare masses $m > m_c$ the system is in the supersymmetric phase
and we observe perfect mass degeneracy between the lowest fermion and
boson mass already at finite lattice spacing. In the supersymmetry
broken phase $m < m_c$ the masses split up and the degeneracy is
lifted. In this phase the masses can also be determined in the
fermionic sector
$Z_{\mathcal{L}_{01}}+Z_{\mathcal{L}_{10}}+Z_{\mathcal{L}_{11}}$ and
we find the same values within our numerical accuracy.  Further
excited states can be obtained by employing the operators $\mathcal{O}
= \psi\phi$ and $\phi^2$ which in the $\mathbb{Z}_2^\chi$ symmetric
phase do not mix with the above operators $\mathcal{O} = \psi$ and
$\phi$, respectively. The result for $\mathcal{O} = \phi^2$ is also
displayed in FIG.~\ref{fig_massdegeneracyfinitespacing}.  In the inset
we show a zoom of the mass $m_\psi^{(0)}$ and the amplitude
$A_\psi^{(0)}$ of the lowest fermionic state. When approaching the
phase transition in the supersymmetry broken phase, the amplitude
decreases and vanishes at the critical point $m_c$, i.e., the particle
decouples from the system when entering the supersymmetric phase. The
mass is by an order of magnitude smaller than the next-to-lowest mass
and the investigation of the finite volume corrections shows that the
data for $m_\psi^{(0)}$ is compatible with zero in the infinite volume
limit, even at finite lattice spacing. Therefore the massless
fermionic mode can be identified with the Goldstino particle
\cite{Salam:1974zb} which mediates the tunnelling between the
fermionic and bosonic vacua and is expected to emerge in the phase
with spontaneously broken supersymmetry. It is quite astonishing
though, that the Goldstino mode materializes so clearly already at
rather coarse lattice spacing.  Finally, from the behavior of the
lowest fermion mass in the supersymmetric phase we can determine the
critical exponent $\nu$ related to the divergence of the fermionic
correlation length. Using $m_\psi^{0}\propto (m-m_c)^{\nu}$ we obtain
$\nu = 0.45 \pm 0.03$ indicating a universality class different from
the Ising one.

\begin{figure}[!tb]
\begin{center}
\includegraphics[scale=0.31]{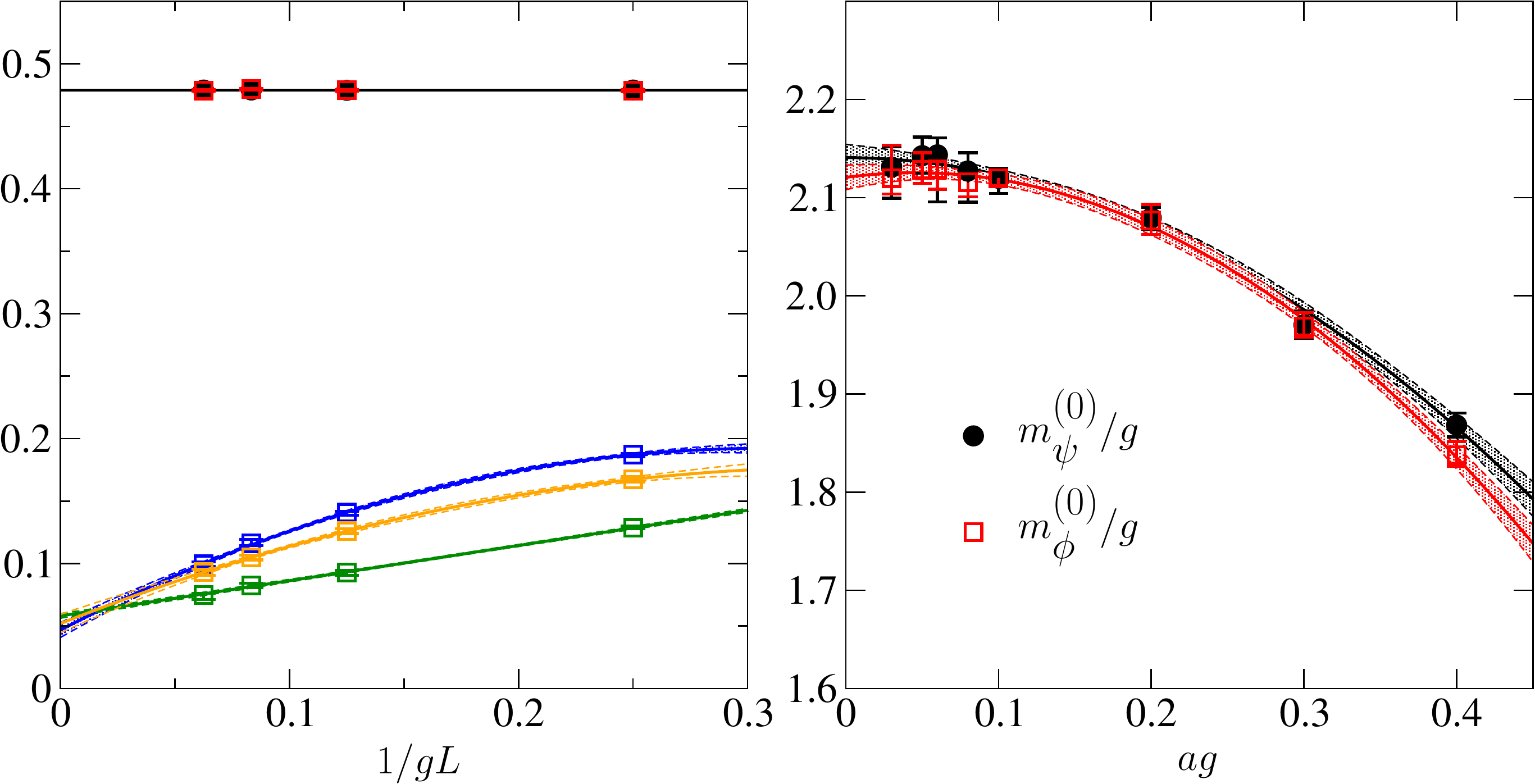}
\caption{{\it Left:} Infinite volume extrapolation of $m^{(0)}_{\phi}$
  at fixed lattice spacing $ag=0.25$ for three bare masses $am=0.30,
  0.18$ and $0.02$ (from bottom up) in the supersymmetry broken phase
  and $m^{(0)}_{\phi}, m^{(0)}_{\psi}$ for one bare mass $am=0.70$ in
  the supersymmetric phase (top). {\it Right:} Continuum extrapolation
  of $m^{(0)}_{\phi}/g$ and $m^{(0)}_{\psi}/g$ at fixed volume $gL=8$
  and renormalised coupling $1/f^R=3$ in the supersymmetric phase.  }
\label{fig_massesthermodyn}
\end{center}
\end{figure}

Next we consider the infinite volume extrapolation of the lowest boson
mass $m^{(0)}_\phi$ in the supersymmetry broken phase for three
different bare masses $am=0.02, 0.18$ and $0.30$ at fixed lattice
spacing in the left panel of FIG.~\ref{fig_massesthermodyn}.
The data clearly indicates a finite boson mass which in the infinite
volume limit is independent of $m$. For comparison we also show the
extrapolation of $m^{(0)}_{\phi}$ and $m^{(0)}_{\psi}$ at a bare mass
$am=0.70$ in the supersymmetric phase where we observe negligible
finite volume effects.  Hence in this phase the continuum
extrapolation of the lowest fermion and boson mass can be done at
fixed physical volume $gL$ and constant renormalised coupling $f^R$.
This is done in the right panel of FIG.~\ref{fig_massesthermodyn}
where we show the data for $gL=8$ and $1/f_R=3$ at different lattice
spacings $ag$. A quadratic plus linear function of $ag$ allows for a
good parametrization of the finite lattice spacing effects. It is
surprising to see that the mass degeneracy holds up even at coarse
lattice spacings where the lattice artifacts are rather strong,
e.g.~$\sim 14$\% at $ag=0.4$.

\begin{figure}[!tb]
\begin{center}
\includegraphics[scale=0.325]{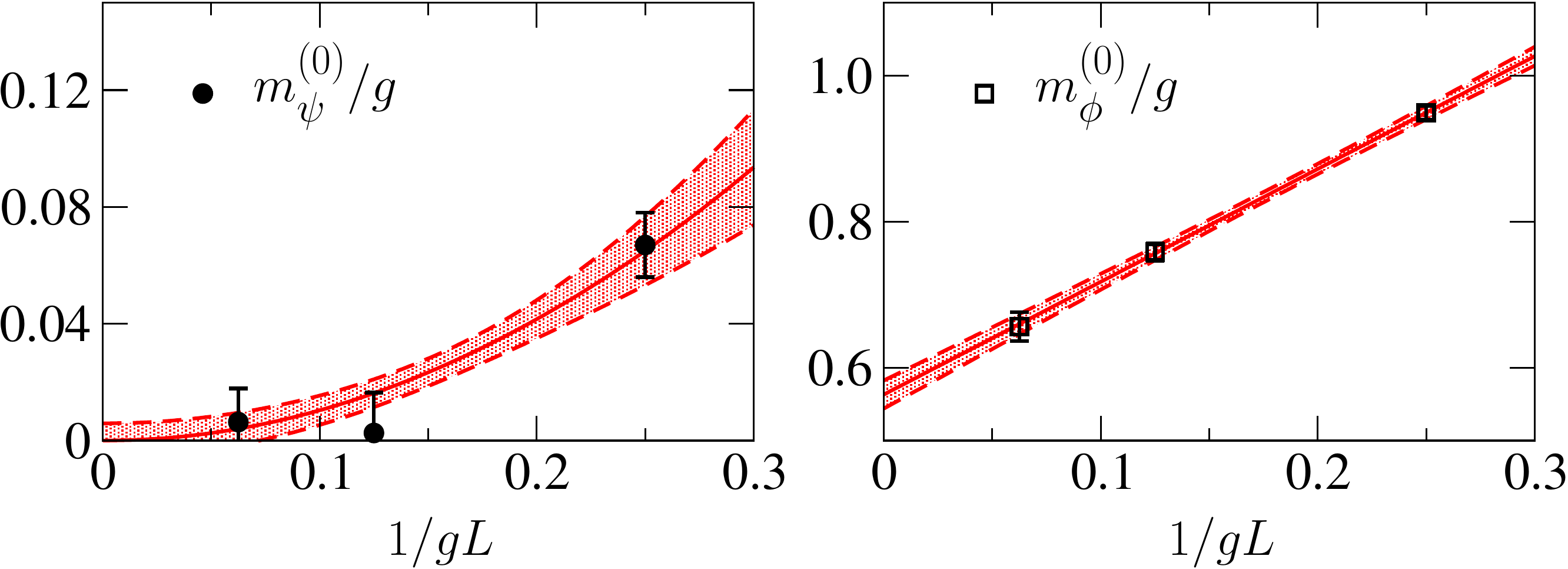}
\caption{Infinite volume extrapolation of
  $m_\psi^{(0)}$ and $m_\phi^{(0)}$ in the supersymmetry broken
  phase at $1/f^R=1.2$.}
\label{fig_degeneracyfixedF}
\end{center}
\end{figure}
Finally, going back to the supersymmetry broken phase we consider the
infinite volume limit of $m^{(0)}_\phi$ and $m_\psi^{(0)}$ in the
continuum at fixed $1/f^R=1.2$ in FIG.~\ref{fig_degeneracyfixedF}. We
find a finite boson mass accompanied by the vanishing fermion mass of
the Goldstino.

\section{Conclusion}

We have established the fermion loop formulation for the
two-dimensional $\mathcal{N}=1$ Wess-Zumino model which allows
efficient simulations with a worm algorithm by avoiding the fermion
sign problem generically appearing in the phase with spontaneously
broken supersymmetry due to the vanishing Witten index. We clearly
observe a $\mathbb{Z}_2^\chi$ symmetric, supersymmetry broken phase
where the bosonic and fermionic vacua (groundstates) are degenerate
and a $\mathbb{Z}_2^\chi$ broken, supersymmetric phase where one of
the two groundstates is spontaneously selected in the infinite volume
limit.  This confirms the expected symmetry breaking pattern and the
corresponding vacuum structure. The phase transition separating those
two phases can be analysed using different observables in the infinite
volume limit and our calculations at several lattice spacings provides
a precise nonperturbative determination of the renormalised critical
coupling in the continuum limit.

Concerning the mass spectrum we observe degenerate boson and fermion
masses in the supersymmetric $\mathbb{Z}_2^\chi$ broken phase,
surprisingly even at finite and rather coarse lattice spacing.  In the
$\mathbb{Z}_2^\chi$ symmetric, supersymmetry broken phase the
nondegeneracy of the lowest few bosonic and fermionic masses can also
be accurately resolved due to the efficient algorithm employed.  The
mass of the lowest fermionic state is compatible with zero in the
thermodynamic and continuum limit, allowing us to identify it with the
expected massless Goldstino mode.

\bibliographystyle{apsrev4-1} 
\bibliography{ssb2dN1WZm}

\end{document}